\documentclass[11pt,twoside]{article}
\usepackage{asp2010}

\resetcounters

\begin{document}

\title{Recent Progress of Cepheid Research at National Central University: From {\it Spitzer} to {\it Kepler}}
\author{Chow-Choong Ngeow}
\affil{National Central University, 320 Jhongli City, Taoyuan Country, Taiwan}

\begin{abstract}

In this presentation I summarize recent work on Cepheid research carried out at the National Central University. The mid-infrared period-luminosity (P-L) relations for Cepheids are important in the {\it James Webb Space Telescope} era for distance scale work, as the relations have potential to derive the Hubble constant within $\sim2$\% accuracy -- a critical constraint in the precision cosmology. Consequently, we have derived the mid-infrared P-L relations for Cepheids in the Large and Small Magellanic Clouds, using archival data from the {\it Spitzer Space Telescope}. {\it Kepler Space Telescope} is a NASA mission to search for Earth-size and larger planets around Sun-like stars, by observing continuously the stars in a dedicated patch of the sky. As a result, the almost un-interrupted observation is also used for stellar variability and asteroseismological study. However, {\it Kepler} observations are carried out with a single broad-band filter, hence ground-based follow-up observation needed to complement {\it Kepler} light curves to fully characterize the properties of the target stars. Here I present the ground-based optical follow-up observations for two Cepheid candidates located within the {\it Kepler's} field-of-view. Together with {\it Kepler} light curves, our ground-based data rule out V2279 Cyg being a Cepheid. Hence V1154 Cyg is the only Cepheid in the {\it Kepler's} field.  

\end{abstract}


\section{Introduction}

Classical Cepheid variables (hereafter Cepheids) are post main-sequence yellow supergiants. Cepheid masses range from $\sim3M_{\odot}$ to $\sim11M_{\odot}$ with surface temperatures between $\sim5000K$ to $\sim6500K$. Cepheids obey the period-mean density relation, hence their pulsating periods ($1<P<100$, where $P$ is period in days) are well correlated with luminosity. This is known as the period-luminosity (P-L) relation, also referred as the Leavitt Law, commonly written in the form of $M_{\lambda}=a_{\lambda}\log(P)+b_{\lambda}$ or $m_{\lambda}=a_{\lambda}\log(P)+b_{\lambda}$. Since Cepheids are evolved pulsating stars, they are ideal laboratories for testing the stellar pulsation and evolution theories. The Cepheid P-L relation is an important rung in the cosmic distance scale ladder that can be used to calibrate a host of secondary distance indicators (for examples, the Tully-Fisher relation, peak brightness of Type Ia supernovae and surface brightness fluctuations). These secondary distance indicators, located well within the Hubble-flow, can then be used to measure one of the most important parameter in modern cosmology -- the Hubble constant. 

In this proceeding, I will present some recent progress on Cepheid research carried out at the National Central University (NCU), that focused on the following two topics: (a) derivation of the mid-infrared Cepheid P-L relations based on archival data from {\it Spitzer} observations; and (b) investigation of potential Cepheids located within the {\it Kepler} field using ground-based follow-up data.

\section{Derivation of the Mid-Infrared Cepheid P-L Relation Using {\it Spitzer} Archival Data}

In the era of precision cosmology, it is desirable to independently derive an accurate Hubble constant, This is because degeneracy exists in CMB (cosmic microwave background) anisotropies between the flatness of the Universe and the Hubble constant. Therefore, as stated in \citet{hu2005}:

\begin{quote}
``... to test the cosmological constant hypothesis and measure the equation of state of the dark energy at $z \sim 0.4$ - $0.5$, the best complement to current and future CMB measurements is a measurement of the Hubble constant that is accurate at the few percent level.''
\end{quote}

\noindent Furthermore, Figures 23 and 24 from \citet{macri2006} clearly illustrate the improved precision of measured dark energy parameters when the accuracy of Hubble constant is increased. Current systematic errors of Hubble constant measurements based on optical data, using the Cepheid P-L relation as first rung of the distance scale ladder, is $\sim5$\% to $\sim10$\% \citep[see the review given in][]{freedman2010}. In the near future, it is possible to reduce the systematic error of Hubble constant to $\sim2$\% from mid-infrared observations from the {\it Spitzer} and/or next generation {\it James Webb Space Telescope (JWST)} \citep{freedman2010}, taking huge advantage of the fact that extinction is negligible in the mid-infrared \citep{freedman2008,ngeow2008,ngeow2009,freedman2010}. The first step toward this goal is to derive and calibrate the Cepheid P-L relation in the mid-infrared.

\subsection{The IRAC Band P-L Relations}

Prior to 2008, the longest available wavelength for the Cepheid P-L relation is in the $K$ band. In 2007, archival data from SAGE (Surveying the Agents of a Galaxy's Evolution) was released, where SAGE is one of the {\it Spitzer} Legacy programs that map out the Large Magellanic Cloud (LMC) in {\it Spitzer} IRAC bands\footnote{The IRAC bands include $3.6\mu \mathrm{m}$, $4.5\mu \mathrm{m}$, $5.8\mu \mathrm{m}$ and $8.0\mu \mathrm{m}$ bands.} \citep[and MIPS bands, see][]{meixner2006}. Hence, mid-infrared Cepheid P-L relations can be derived by matching the known LMC Cepheids to the SAGE catalog -- this has resulted two papers published in 2008 \citep[][]{freedman2008,ngeow2008}. Both papers utilized the SAGE Epoch 1 data matched to different samples of LMC Cepheids: \citet{freedman2008} matched to $\sim70$ Cepheids from \citet{persson2004} while \citet{ngeow2008} used the OGLE-II Catalog \citep[Optical Gravitational Lensing Experiment,][]{udalski1999} that includes $\sim600$ LMC Cepheids.  

After the release of SAGE Epoch 1 and 2 data, the IRAC band LMC P-L relations have been updated in \citet{madore2009} and \citet{ngeow2009} by averaging the two epochs photometry. \citet{ngeow2009} also included the $\sim1800$ LMC Cepheids from the latest OGLE-III Catalog \citep[][with $\sim3 \times$ more Cepheids than OGLE-II Catalog]{soszynski2008}. In addition to LMC, IRAC band P-L relations have also been derived for SMC Cepheids \citep[see][for details]{ngeow2010} using the publicly available SAGE-SMC data. Finally, IRAC band P-L relations for a small number of Galactic Cepheids that possess independent distance measurements have been presented by \citet{marengo2010}. All together, there are six sets of IRAC band P-L relations for Cepheids in our Galaxy and in Magellanic Clouds. Slopes of these P-L relations are summarized in Figure \ref{fig2}. 

\articlefiguretwo{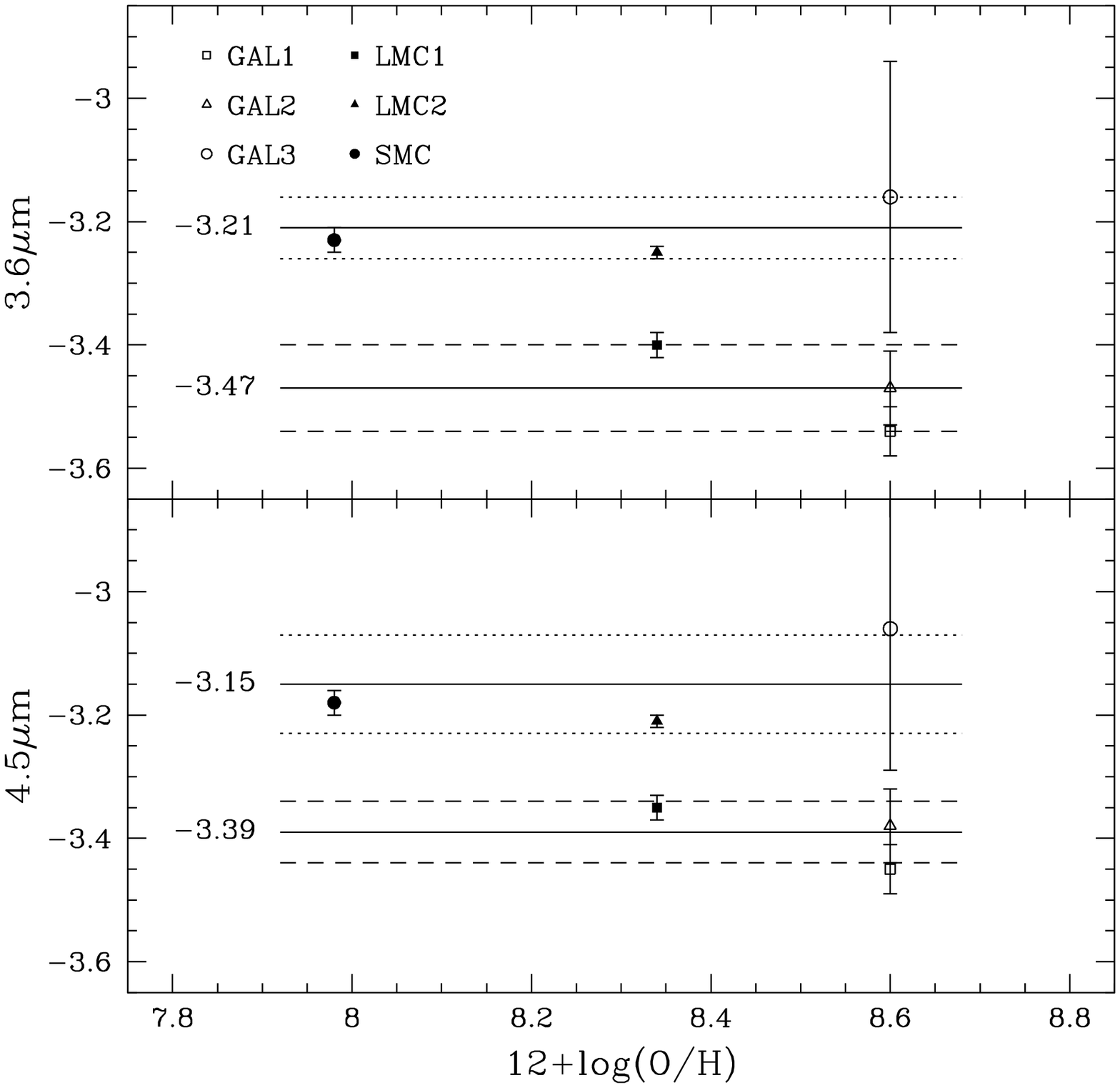}{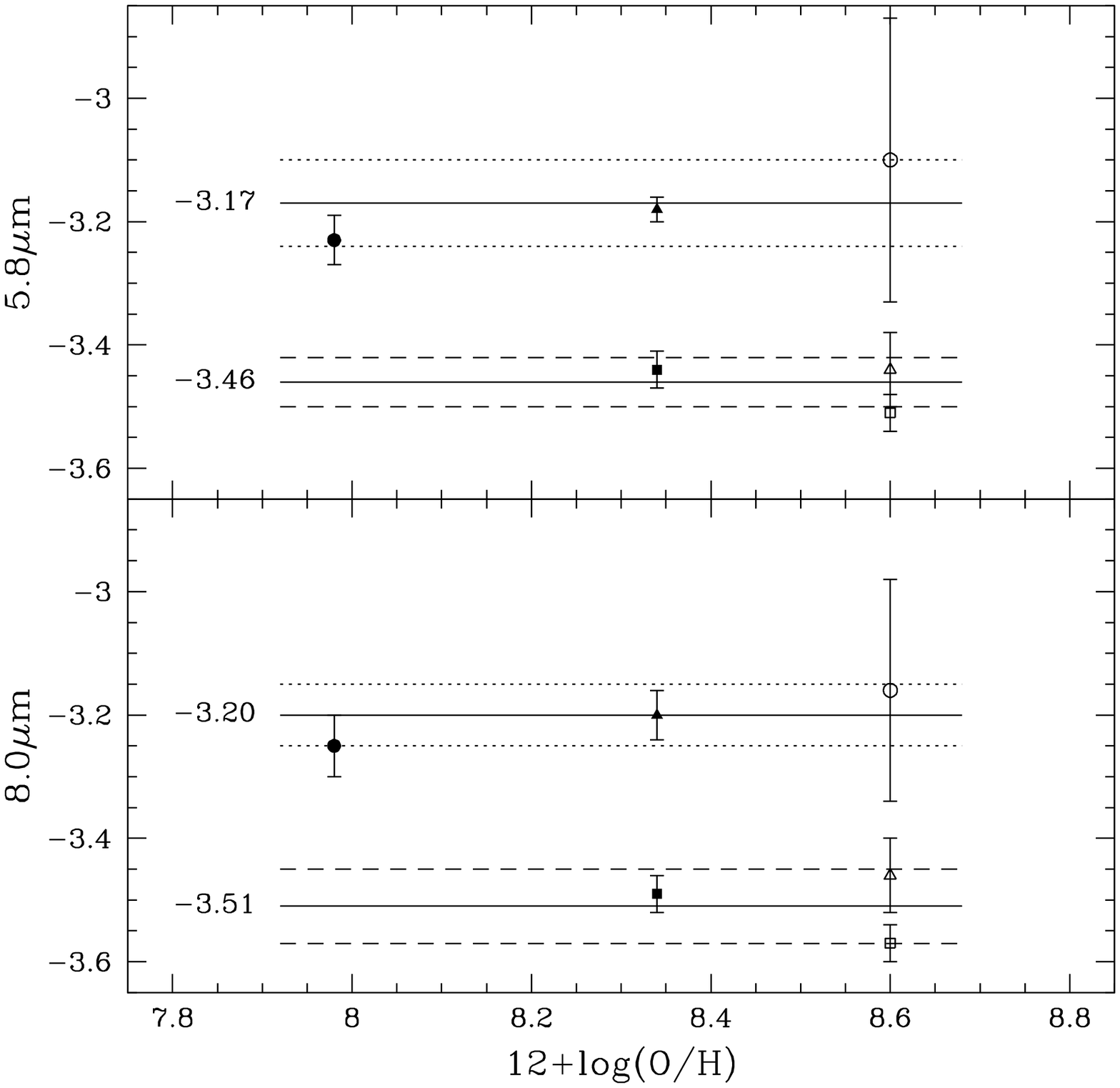}{fig2}{Slopes of the empirical IRAC band P-L Relations from three galaxies. GAL1 and GAL2 were derived from Galactic Cepheids with infrared surface brightness distances using the ``old'' and ``new'' $p$-factor, respectively; while GAL3 was based on $8$ Galactic Cepheids with {\it HST} parallax measurements \citep[for more details, see][]{marengo2010}. LMC1 and LMC2 are the empirical LMC P-L slopes from \citet{madore2009} and \citet{ngeow2009}, respectively. SMC P-L slopes are adopted from \citet{ngeow2010}. These P-L slopes can be grouped into the ``shallow'' slopes and ``steep'' slopes. The Horizontal dashed lines represent the averaged slopes in these two groups, with the values given on the left of these lines. The dashed and dotted lines are the $1\sigma$ boundary of the averaged values for the slopes in these two groups, where $\sigma$ is the standard deviation of the mean values.}

As mentioned in \citet{ngeow2010}, these six sets of P-L slopes can be grouped into two groups: one group with ``shallow'' slopes around $-3.18$, and another group with ``steeper'' slopes around $-3.46$. The expected IRAC band P-L slope can be calculated from $L_{\lambda}=4\pi R^2 B_{\lambda}(T)$, hence $M_{\mathrm{IRAC}}=-5\times a_R\log(P)+a_T\log(P)+\mathrm{constant}$ \citep[where $a_R=0.68$ is the slope of period-radius relation taken from][]{gieren1999}. If assumed $B_{\lambda}(T)$ is constant at long wavelength, then the expected P-L slops is $-5\times 0.68=-3.40$. On the other hand, due to Rayleigh-Jean approximation, $B_{\lambda}(T)\propto T$ at the IRAC band wavelength, then $a_T\sim0.18$ estimated from color-temperature conversion \citep[for details, see][]{neilson2010}, the expected P-L slope becomes $-5\times 0.68 + 0.18 =-3.22$. Both of these expected slopes are close to the averaged slopes in the observed ``steep'' and ``shallow'' groups. Finally, these empirical P-L slopes were compared to the slopes from synthetic P-L relations, based on theoretical pulsational models with varying metallicity, in Figure \ref{fig3}. Details of these synthetic P-L relations will be given elsewhere (Ngeow et al. -- in preparation). The trends of synthetic P-L slopes with $12+\log(O/H)>7.9$ appear to be in agreement to the empirical slopes given in ``shallow'' group.

\articlefigure[scale=0.5]{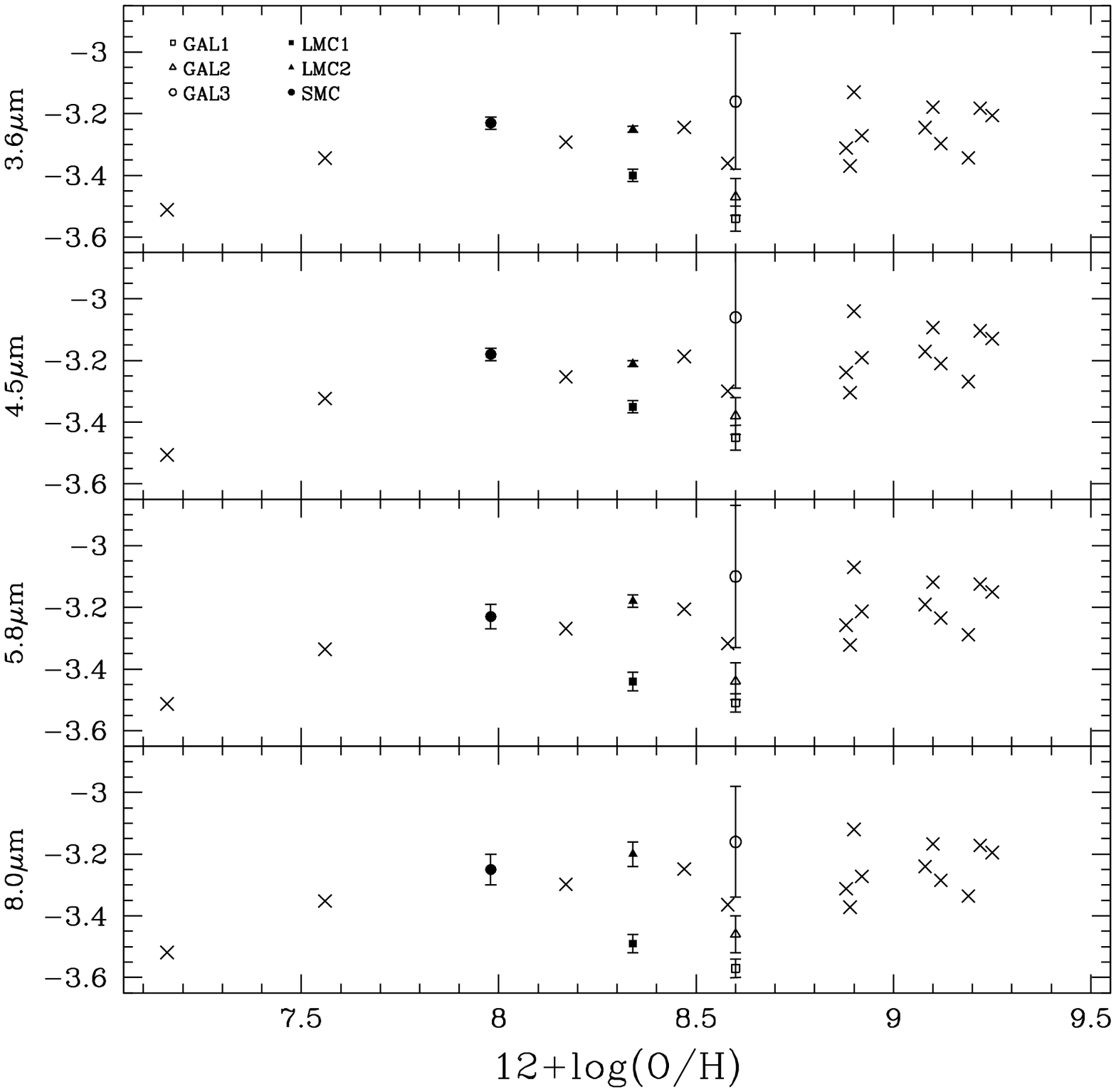}{fig3}{Comparison of the six sets of empirical IRAC band P-L slopes to the synthetic P-L slopes from theoretical models (crosses).}

\section{Ground-Based Light Curves for Cepheids Located Within {\it Kepler} Field-Of-View}

{\it Kepler} is a NASA space mission to search for Earth-like exo-planets using the transit method. Hence, the field-of-view for {\it Kepler}, with $\sim105\mathrm{deg}^2$, is constantly pointing toward $19^\mathrm{h}22^\mathrm{m}$:$+44^o30'$. In addition to searching for exo-planets, light curve data from almost un-interrupted observations from {\it Kepler} are also very useful and valuable for asteroseismic and stellar variability study. Therefore, a consortium called {\it Kepler} Asteroseismic Science Consortium (KASC)\footnote{{\tt http://astro.phys.au.dk/KASC/}} was formed. KASC consists of 14 working groups (WG), and WG7 is dedicated to Cepheids study with {\it Kepler} data. 

Since Cepheids are radially pulsating variable stars with periods longer than a day, it is not necessary to continuously monitor Cepheids in a given night or from different observatories across the globe. These multiple-site observing campaigns, on the other hand, have been practiced for other kinds of variable stars such as short period delta-Scuti stars to investigate, for example, various pulsating modes of these stars. Continuous photometry from {\it Kepler}, however, offers a great opportunity to study the possible non-radial pulsating modes as suggested by theory \citep{mulet2007}, or even stochastically excited modes, in Cepheids. However, {\it Kepler} observations are conducted in a customized broad-band filter, and complementary ground-based multi-color photometric and spectroscopic follow-up observations are needed for the Cepheid candidates located within the {\it Kepler} field-of-view. Detailed investigation of these candidates using both of the {\it Kepler} light curves and ground-based follow-up data can be found in \citet{szabo2011}, only a brief overview will be presented in this proceeding.

\subsection{$BVRI$ Follow-Up Observations for Selected Cepheid Candidates}

Prior to the launch of {\it Kepler}, V1154 Cyg was the only Cepheid located in {\it Kepler's} field, while V2279 Cyg was a strong Cepheid candidate. Several other Cepheid candidates were also selected based on the {\it Kepler} Input Catalog or other variable stars catalogs \citep{szabo2011}. Ground-based $BVRI$ follow-up observations for these Cepheid candidates were carried out at Lulin Observatory (located at central Taiwan) and Tenagra II Observatory (located in Arizona, USA), mostly from March to August 2010. The author is responsible for the observations from these two observatories, where the telescopes and CCD used are listed in Table 1. Figure \ref{fig4} shows the distributions of $FWHM$ in $BVRI$ band images taken from these telescopes. Data reductions were done based on the following steps:

\begin{table}[!ht]
\caption{Characteristics of telescopes and CCD for the $BVRI$ follow-up observations.}
\smallskip
\begin{center}
{\small
\begin{tabular}{ccccc}
\tableline
\noalign{\smallskip}
Tel. Abbreviation  & Observatory & Aperture [m] & CCD & Pixel Scale [''/pix]\\ 
\noalign{\smallskip}
\tableline
\noalign{\smallskip}
LOT & Lulin Obs. & 1.0 & PI-1300B & 0.52 \\
SLT & Lulin Obs. & 0.4 & Apogee U9000 & 0.99 \\
TNG & Tenagra II Obs. & 0.8 & STIe & 0.86 \\
\noalign{\smallskip}
\tableline
\end{tabular}
}
\end{center}
\end{table}

\articlefiguretwo{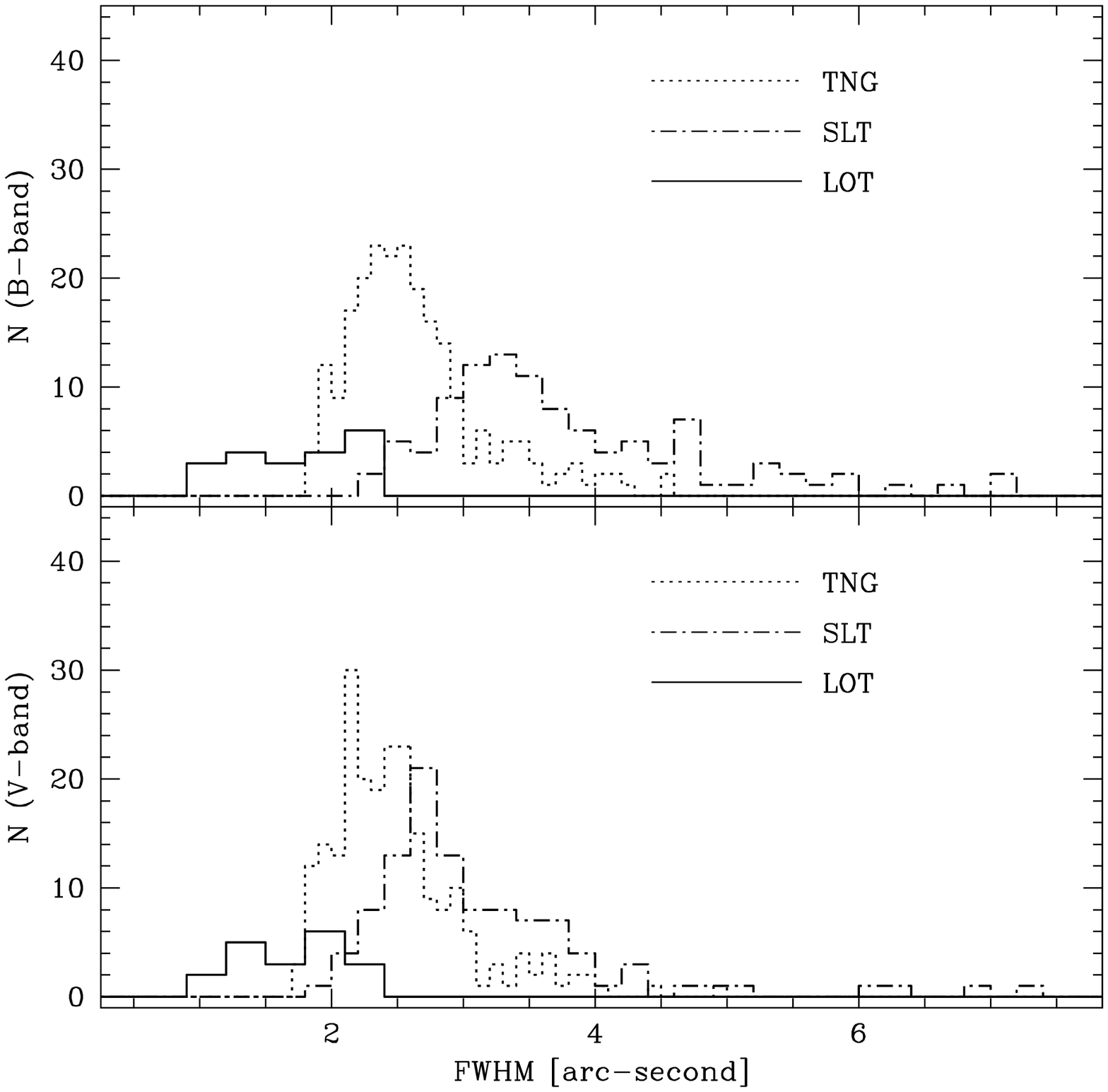}{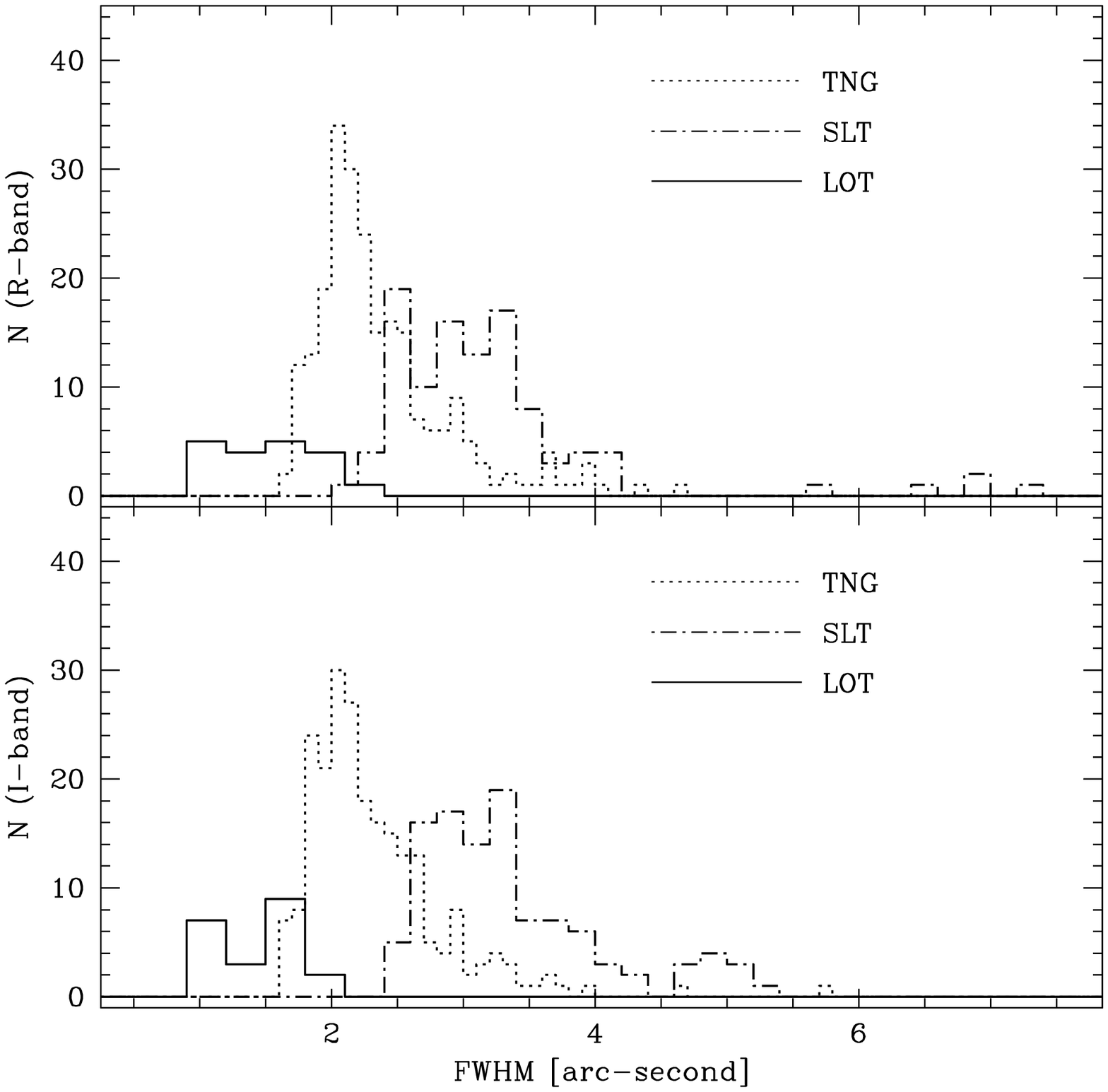}{fig4}{Histograms of the $FWHM$ distributions in $BVRI$ bands, separated by the telescopes used.}

\begin{enumerate}
\item Images were bias-subtracted, dark-subtracted and flat-fielded using $IRAF$\footnote{$IRAF$ is distributed by the National Optical Astronomy Observatories, which are operated by the Association of Universities for Research in Astronomy, Inc., under cooperative agreement with the National Science Foundation.}. Fringe corrections were applied to the $I$ band images from LOT.
\item Astrometric corrections were performed using {\tt astrometry.net} \citep{lang2010} and {\tt SCAMP} \citep{bertin2006}.
\item Instrumental magnitudes were extracted from the images using {\tt SExtractor} \citep{bertin1996}. These magnitudes were based on {\tt MAG\_AUTO} aperture photometry.
\item Observations from Tenagra II Observatory on 12 May 2010 included the Landolt standard fields \citep{landolt2009}. Instrumental magnitudes from this night were calibrated to the standard magnitudes.
\item Photometry from other nights were relatively calibrated using the calibrated magnitudes from 12 May 2010. 
\end{enumerate}

\noindent The resulted $BVRI$ light curves for two of the candidates, V1154 Cyg and V2279 Cyg, are presented in Figure \ref{fig5}. Additional data were available from Sonoita Research Observatory\footnote{Observations and the data reduction from Sonoita Research Observatory have been taken care by A. Henden.} \citep{szabo2011}.

\articlefiguretwo{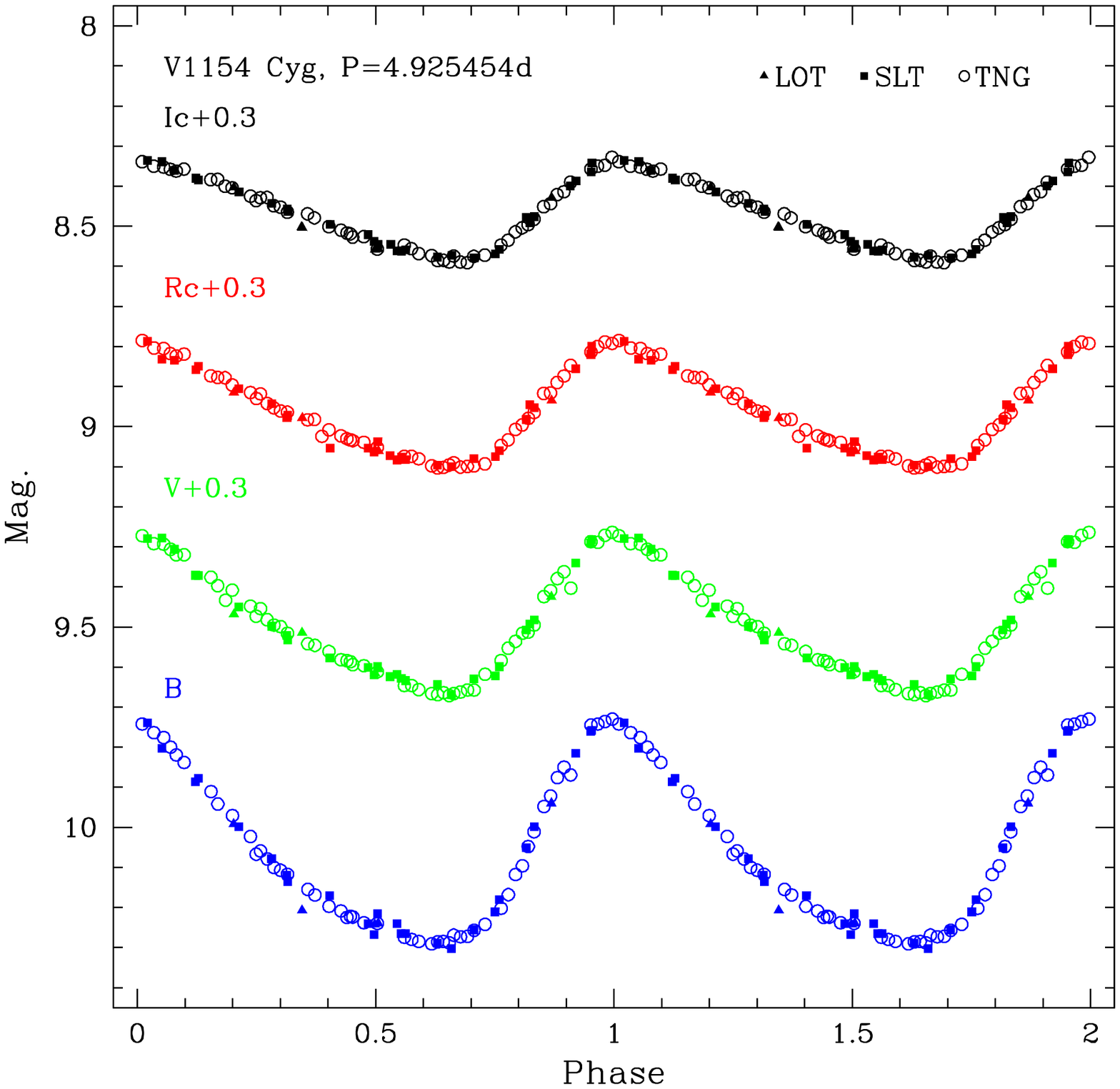}{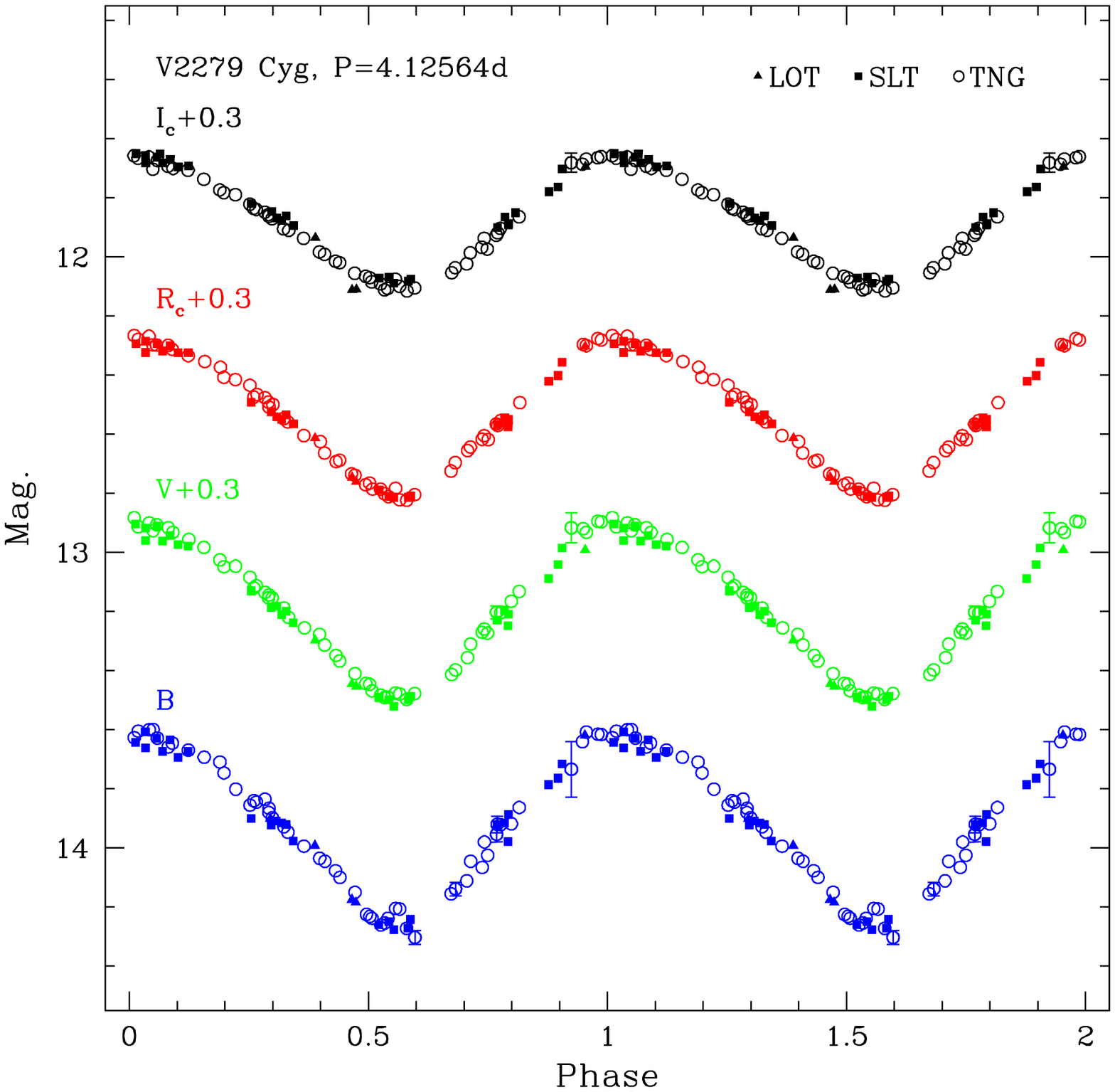}{fig5}{Calibrated $BVRI$ light curves for V1154 Cyg (left panel) and V2279 Cyg (right panel). Each light curves contain more than $70$ data points.}

Light curves of V1154 Cyg from {\it Kepler} observations resemble a typical sawtooth shape of a Cepheid's light curve \citep[see Figure 4 in][]{szabo2011}. Frequency analysis of almost continuous {\it Kepler} light curves for this variable showing a strong peak at the fundamental frequency, with other detectable harmonics in the spectrum \citep[see Figure 12 in][]{szabo2011}. This suggests that V1154 Cyg pulsates radially in a regular fashion, without any non-radial or stochastic modes. The $BVRI$ light curves of V1154 Cyg also strongly support the Cepheid nature of this variable. Fourier parameters ($R_{i1}$ and $\phi_{i1}$), based on the $V$ band light curve, of this Cepheid also fall within the distributions defined by Galactic Cepheids. The phase lag from $V$ band light curve and radial velocity curve confirms that V1154 Cyg is a fundamental mode Cepheid.

Even though the $BVRI$ band light curves for V2279 Cyg mimic the light curve of a first overtone Cepheids, Fourier analysis of this variable revealed that this star is not a Cepheid. Furthermore, flares show up in {\it Kepler} light curves, and the light curve morphology is in agreement with rotational modulation. Additional spectroscopic follow-up observations confirmed that this variable is not a Cepheid. Cepheid nature of other candidates has been ruled out based on {\it Kepler} light curves and/or $BVRI$ light curves, and V1154 Cyg remains the only Cepheid located within the {\it Kepler's} field. 

\section{Conclusion}

I have presented some recent work on Cepheid research that carried out at NCU. In summary: 

\begin{itemize}
\item Mid-infrared P-L relations will be important in near future  and in {\it JWST} era with the potential to reduce the systematic error of Hubble constant. Consequently, P-L relations based on Cepheids in Magellanic Clouds have been derived using the archival {\it Spitzer} data.

\item Almost continuous observation from {\it Kepler} is very valuable for stellar variability and pulsation study. However, ground-based follow-up observations are needed to complement {\it Kepler} data to fully investigate the variable stars in the {\it Kepler's} field. Together with {\it Kepler} light curves, the ground-based follow-up observations for Cepheid candidates confirmed that V1154 Cyg is the only Cepheid located within {\it Kepler's} field.
\end{itemize}

\acknowledgements 

CCN would like to thank the support from National Science Council under the contract NSC98-2112-M-008-013-MY3. We would also like to thank the collaboration and discussion with Shashi Kanbur, Hilding Neilson, Marcella Marconi, Ilaria Musella, Michele Cignoni, Robert Szabo, Laszlo Szabados, Arne Henden and the {\it Kepler} Cepheid Working Group. This work is based [in part] on observations made with the {\it Spitzer Space Telescope}, which is operated by the Jet Propulsion Laboratory, California Institute of Technology under a contract with NASA. Funding for the {\it Kepler} Mission is provided by NASA’s Science Mission Directorate. We acknowledge assistance of the queue observers, Chi-Sheng Lin and Hsiang-Yao Hsiao, from Lulin Observatory, which is operated by the National Central University of Taiwan, and is funded partially by the grant NSC97-2112-M-008-005-MY3. We also acknowledge the variable star observations from the AAVSO International Database contributed by observers worldwide and used in this research.

\bibliography{ngeow_prcsa2011}

\end{document}